\documentclass[onecolumn,useAMS,usenatbib]{mn2e}

\usepackage{color}
\usepackage{amsmath}
\usepackage{graphicx}
\usepackage{amssymb}
\usepackage{psfrag}

\newcommand{\beq}{\begin{equation}}
\newcommand{\eeq}{\end{equation}}
\newcommand{\beqy}{\begin{eqnarray}}
\newcommand{\eeqy}{\end{eqnarray}}

\newcommand{\Bav}{\bar{B}}

\newcommand{\brac}[1]{\left( {#1} \right)}

\newcommand{\be}{{\bf{e}}}
\newcommand{\br}{{\bf{r}}}
\newcommand{\bv}{{\bf{v}}}
\newcommand{\dvmu}{\frac{1}{4\pi}}
\newcommand{\bOm}{{\boldsymbol{\Omega}}}
\newcommand{\bB}{{\bf B}}
\newcommand{\pd}[2]{\frac{\partial{#1}}{\partial{#2}}}
\newcommand{\curl}{\nabla\times}
\renewcommand{\div}{\nabla\cdot}

\newcommand{\dP}{\delta P}

\newcommand{\drh}{\delta\rho}
\newcommand{\dB}{\delta\bB}
\newcommand{\boldf}{{\bf f}}

\newcommand{\rmd}{{\mathrm d}}

\newcommand{\clP}{\mathcal{P}}

\title[Oscillations and instabilities in NSs]{Oscillations and
  instabilities in neutron stars with poloidal magnetic fields}
\author[S. K. Lander and D. I. Jones]
       {S. K. Lander\thanks{skl@soton.ac.uk} and D. I. Jones\\
School of Mathematics, University of Southampton, Southampton SO17 1BJ}
\begin{document}


\pagerange{\pageref{firstpage}--\pageref{lastpage}} \pubyear{0000}

\maketitle

\label{firstpage}

\begin{abstract}
We study the time evolution of non-axisymmetric linear perturbations
of a rotating magnetised neutron star, whose magnetic field is
purely poloidal. The background stellar configurations
are generated self-consistently, with multipolar field
configurations and allowing for distortions to the
density distribution from rotational and magnetic forces. The
perturbations split into two symmetry classes, with perturbations in
one class being dominated by an instability
generic to poloidal fields, which is localised around the `neutral
line' where the background field vanishes. Rotation acts to reduce the
effect of this instability. Perturbations in the other symmetry class do
not suffer this instability and in this case we are able to resolve
Alfv\'en oscillations, whose restoring force is the magnetic
field; this is the first study of non-axisymmetric Alfv\'en modes of a
star with a poloidal field. We find no evidence that these modes form a
continuum. In a rotating magnetised star we find 
that there are no pure Alfv\'en modes or pure inertial modes, but hybrids
of these. We discuss the nature of magnetic instabilities and
oscillations in magnetars and pulsars, finding the dominant Alfv\'en
mode from our simulations has a frequency comparable with observed
magnetar QPOs.
\end{abstract}


\section{Introduction}

Neutron stars are notable for the extreme strength of their magnetic
fields, with surface fields reaching $\sim 10^{15}$ G for magnetars
and interior fields perhaps being an order of magnitude stronger
still. We expect many aspects of neutron star (NS) physics to be influenced
by their magnetic fields, but we still have limited understanding of
the actual structure of these fields; there are still open questions
concerning their strength in the stellar interior, the relative
proportions of poloidal and toroidal components and the possible effect of
superconductivity (among others). One particular motivation for
improving modelling of NS magnetic fields is the observation of
quasi-periodic oscillations (QPOs) in the aftermath of giant flares from
magnetars; these provide the first direct evidence of NS oscillations
and give us a potential probe of the interior physics of these stars.

One way to build up an improved understanding of NS magnetic fields is
to explore the equilibria and dynamics of a simplified NS model. We
choose to study a non-relativistic fluid star, but allowing for the
effects of rotation and a magnetic field. In \citet{landerjones} we
explored NS structures within this model, generating
self-consistent equilibrium solutions for stars
with purely poloidal, purely toroidal and mixed poloidal-toroidal
magnetic fields. To understand the stability and oscillations of these
stars we perform time evolutions of perturbations, using our NS
equilibria as background configurations. In \citet{tor_mode} we studied
the oscillation spectrum of NSs with purely toroidal magnetic fields,
whilst \citet{tor_instab} presented results for toroidal-field
instabilities. This paper discusses oscillations and instabilities of
poloidal-field NSs, whilst in future we hope to complete this study of
magnetised NSs by exploring mixed-field configurations.

We begin by discussing the governing equations for our NS model, both
for the background equilibria and for the time-evolution of the
perturbations. We find that perturbations in one symmetry class are unstable in
a region where the background field vanishes (consistent with the
analytic work of, e.g., \citet{wright} and \citet{taylerpol}), with
this unstable behaviour being reduced by the effect of rotation. We
quantify the effect of magnetic field 
strength and rotation on the instability's growth rate. Perturbations
in the other symmetry class, by contrast, appear to evolve stably; this
confirms a prediction from \citet{taylerpol} about the nature of the
dominant poloidal-field instability. For these stable perturbations we
are able to resolve many Alfv\'en oscillation periods and 
extract their mode frequencies; this is the first study of
non-axisymmetric Alfv\'en modes in a poloidal-field star,
complementing a number of recent studies on axisymmetric oscillations
\citep{sotani_ax,sotani_pol,cerda,colaiuda}. We next turn to the
oscillations of rotating magnetised stars, 
showing that these are a hybrid of inertial and Alfv\'en
modes. We conclude by discussing our work in the context of magnetars and
pulsars and discuss the possible proportions of poloidal and toroidal
components required for stability.

\section{Governing equations and numerics}

We begin by describing the equations governing our NS model, both for
the axisymmetric stationary background configuration and the
non-axisymmetric perturbations. A fuller account of the numerical
methods used in the code and its performance may be found in
\citet{tor_mode}; these are only described briefly here.

We model a neutron star as a self-gravitating, rotating, magnetised
polytropic fluid with perfect conductivity, in Newtonian gravity. We
wish to study linear 
perturbations of this star; for this our governing equations consist
of a set of stationary background equations and a set of
equations describing the time evolution of the perturbations. The
background configuration has a purely poloidal magnetic field
$\bB_0$ and may be (rigidly) rotating:
\beq \label{backeuler}
0 = -\nabla P_0 - \rho_0\nabla\Phi_0
    - \rho_0\bOm\times(\bOm\times\br) + \dvmu(\curl\bB_0)\times\bB_0,
\eeq
\beq
\nabla^2\Phi_0 = 4\pi G\rho_0,
\eeq
\beq
P_0=k\rho_0^\gamma,
\eeq
where $P$ is stellar pressure, $\rho$ density, $\Phi$ gravitational
potential, $G$ gravitational constant and $\bOm$ angular velocity;
$0$-subscripts denote background quantities. Finally, we will take
$\gamma=2$ throughout this study, as a rough approximation to a
neutron star equation of state.

Many studies of poloidal-field oscillations assume a dipolar
field configuration, given by some simple analytic expression. In
constrast to these, we solve for the field and fluid together, using a
non-linear iterative procedure. The result is a self-consistent field
configuration composed of a sum of different multipolar
contributions. These higher multipoles are more significant
  in more distorted stars, providing small corrections to nonrotating
  and highly magnetised stars and becoming comparable with the dipolar
  field in rapidly rotating stars. Using the magnetic vector potential
${\bf A}$ defined through $\bB=\curl{\bf A}$, one may show that the
magnetic vector potential (in spherical polar coordinates) satisfies
the equation
\beq
\nabla^2(A_\phi\sin\phi) = -\kappa\rho_0 r\sin\theta\sin\phi,
\eeq
where $\kappa$ is a constant governing the strength of the
field; a derivation of this is given in \citet{landerjones}. From the
vector potential we obtain the background magnetic field configuration:
\beq
\bB_0=\curl(A_\phi\be_\phi);
\eeq
since $\bB$ is poloidal it is described by a \emph{toroidal} vector
potential. These background equations are solved numerically using an
iterative procedure to find stationary equilibrium configurations, as
detailed in \citet{landerjones} (see also \citet{tomi_eri}). Our
method allows for distortions of 
the star due to both rotational and magnetic forces; for the purely
poloidal fields considered here, both of these forces will act to
make the star oblate.

For the perturbation equations, we work in the rotating frame of the
background and write our equations in terms of the perturbed density
$\drh$, the mass flux $\boldf=\rho_0\bv$ and a magnetic variable
$\bbeta=\rho_0\dB$. We additionally make the Cowling approximation ---
neglecting the perturbed gravitational force --- to avoid the
computational expense of solving the perturbed Poisson equation. Our
perturbations are then governed by seven equations:
\beq \label{euler_magmode}
\rho_0\pd{\boldf}{t}
  = -\gamma P_0\nabla\drh - 2\bOm\times\boldf
    + \brac{\frac{(2-\gamma)\gamma P_0}{\rho_0}\nabla\rho_0
                            - \dvmu(\curl\bB_0)\times\bB_0}\drh
    + \dvmu(\curl\bB_0)\times\bbeta + \dvmu(\curl\bbeta)\times\bB_0
    - \frac{1}{4\pi\rho_0}(\nabla\rho_0\times\bbeta)\times\bB_0,
\eeq
\beq \label{conti_magmode}
\pd{\drh}{t}=-\div\boldf,
\eeq
\beq \label{induc_magmode}
\pd{\bbeta}{t} = \curl(\boldf\times\bB_0)
                 -\frac{\nabla\rho_0}{\rho_0}\times(\boldf\times\bB_0).
\eeq

The axisymmetry of the background allows us to decompose the
perturbation equations in the azimuthal index $m$, thus reducing the
three-dimensional system of equations to a 2D one. It is also
beneficial to isolate perturbations of a particular $m$ when discussing
oscillation modes and instabilities.

\subsection{Boundary conditions}
\label{BCs}

Instead of working with the spherical-polar radial coordinate $r$, we
employ a coordinate $x$ fitted to isopycnic surfaces, which for
nonspherical stars is a function of $r$ and $\theta$. Doing this gives
us a very simple set of boundary conditions at the stellar
surface:
\beq
\boldf(x=R)=\bbeta(x=R)={\bf 0}\ ,\ \dP(x=R)=0.
\eeq
Since we deal with $m\neq 0$ perturbations, we should also enforce a
zero-displacement condition at the centre:
\beq
\boldf(x=0)=\bbeta(x=0)={\bf 0}\ ,\ \dP(x=0)=0.
\eeq
Perturbation variables are also zero at the pole, except for $m=1$
where $f_\theta,f_\phi,\beta_\theta$ and $\beta_\phi$ are nonzero; in
the case of these quantities, it is their $\theta$-derivatives which
vanish.

Finally, it may be shown that perturbations of stars with purely
poloidal background fields form two sets based on shared symmetry
properties: 
$S_1\equiv\{f_r,f_\phi,\drh,\beta_\theta\}$ and 
$S_2\equiv\{f_\theta,\beta_r,\beta_\phi\}$. All the perturbations
within a set have the same equatorial symmetry (say, even/symmetric),
whilst all the perturbations in the other set have the opposite
symmetry (which will then be odd/antisymmetric).
We may enforce these symmetries as an additional set of boundary conditions,
reducing our (2D) numerical domain to one quadrant of a disc.
There are,
therefore, two symmetry classes of solution: one class $\clP^+$ where
elements of $S_1$ are even (with those in $S_2$ odd) and one class
$\clP^-$ where $S_1$-elements are odd (with $S_2$-elements even).

For more details on the boundary conditions summarised here, see
\citet{tor_mode} or \citet{tor_instab}. Finally, we note that whilst
the poloidal background field vanishes on a `neutral line' within the
star (see section \ref{instab_section}), we do not impose vanishing of
the perturbations there.

\subsection{Initial data}
\label{initial_data}

Depending on the initial data chosen and the symmetry class specified
(see above), either `axial-led' or `polar-led' perturbations (using the
terminology of \citet{lock_fried}) will be excited. In particular,
axial-led perturbations are those in the class $\clP^-$ and polar-led
perturbations are in the $\clP^+$ class. For evolutions of 
polar-led perturbations we begin with an initial spherical-harmonic
perturbation in the density: $\drh\sim Y_{lm}(\theta,\phi)$, whilst if
we wish to excite axial-led perturbations we prescribe a `magnetic'
spherical harmonic perturbation in the velocity: $\bv\sim\nabla
Y_{lm}\times{\be_r}$.

Evolving perturbations on a background star with a poloidal field, we
find that axial initial data strongly excites an unstable
mode. Although some oscillatory behaviour can be seen too, the
unstable mode dominates the evolutions and prevents us from extracting
information about axial-led modes. With polar initial data, however,
there is no evidence of an instability, and in this case we are
able to resolve enough Alfv\'en oscillations to study polar-led mode
frequencies. For this reason, all results in the section on
instabilities come from evolutions using axial initial data (and
  the corresponding symmetry class $\clP^-$), whilst
all results in the oscillations section use polar initial data (and
the other symmetry class $\clP^+$).

\subsection{Numerics}

As described above, we need only evolve perturbations on one quadrant
of a disc, by exploiting symmetries of the original 3D system of
equations. We employ a McCormack predictor-corrector algorithm to
evolve the perturbation equations in time. We add an artificial viscosity term
to our equations to damp out numerical instabilities that can emerge
when using finite-difference routines, but we ensure that this is as
small as possible --- otherwise it could unnecessarily damp the
\emph{physical} hydromagnetic instabilities which we aim to study. We
additionally employ a divergence-cleaning routine, since numerical
error in the evolution of the perturbed magnetic field can result in
an unphysical monopolar term: $\div\dB\neq 0$ \citep{dedner}. We
do not use artificial resistivity in this work.

\subsection{Dimensionless and physical quantities}

In our code we nondimensionalise all quantities by dividing them by
the requisite combination of gravitational constant $G$, equatorial
radius $R$ and maximum stellar density $\rho_m$. Some results
presented in this paper are in terms of dimensionless quantities; this
is either because they are frequently used in the literature or
because there is no particular benefit to quoting them in physical
units. Dimensionless quantities are denoted with a hat; for example
$\hat\Omega=\Omega/\sqrt{G\rho_m}$. Note that dimensionless
frequencies (i.e. the rotation $\Omega$ or mode frequency
$\sigma$) are quoted in terms of \emph{radians} rather than
\emph{cycles} per unit code-time.

In all cases where we have redimensionalised our results, this is done
to the same `canonical' neutron star, with a mass of
1.4 solar masses and whose radius would be $10$ km if it were non-rotating
and unmagnetised (i.e. in hydrostatic equilibrium and
spherical). For this canonical star the maximum density is $2.2\times
10^{15}$ g cm${}^{-3}$. Rotation frequencies may be  
redimensionalised using the approximate relation
\beq
\Omega\ [\mathrm{rad s}{}^{-1}]\approx 12100\hat\Omega,
\eeq
with the same relation holding for mode frequencies $\sigma$. The
corresponding rotation frequency $\nu$ in \emph{cycles} per second may be
obtained using
\beq
\nu\ [\mathrm{Hz}]=\frac{\Omega\ [\mathrm{rad s}{}^{-1}]}{2\pi}
                               \approx 1900\hat\Omega
\eeq
--- again, the same relation may be used to obtain mode frequencies in
Hz.

\vspace{-0.2cm}
\section{Instabilities of purely poloidal fields}
\label{instab_section}

\begin{figure}
\begin{center}
\begin{minipage}[c]{0.5\linewidth}
\psfrag{open}{open field lines}
\psfrag{closed}{closed field lines}
\psfrag{neutral}{neutral line}
\includegraphics[width=\linewidth]{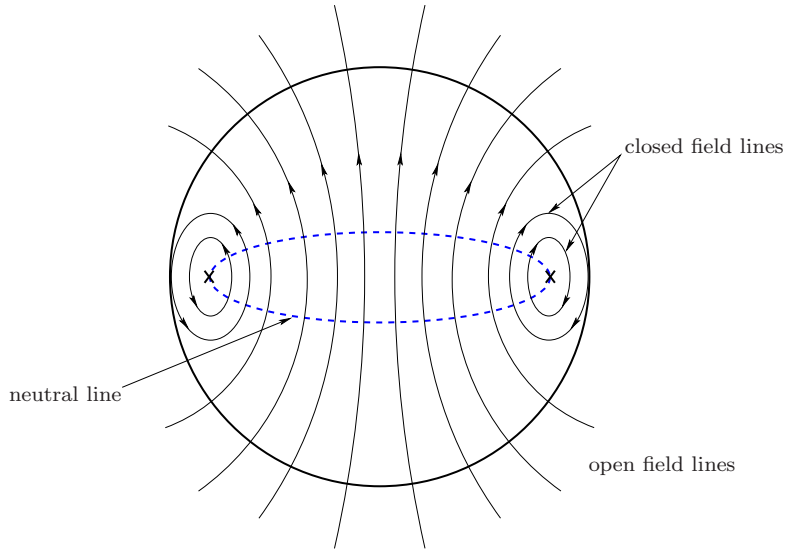}
\end{minipage}
\caption{\label{fieldlines}
         Schematic field-line geometry for a poloidal magnetic
         field. Around the star's 
         symmetry axis the field lines are `open', in the sense that
         they extend outside the star; near the equatorial surface
         there is a region of field lines which close within the
         star. At the centre of these closed field lines is the
         neutral line (represented by the crosses in 2D, with the
         dashed line showing its circular form in the 3D star); along
         this line the magnetic field strength drops to zero.}
\end{center}
\end{figure}

The first indications that purely poloidal fields could generically
suffer hydromagnetic instabilities appeared in the studies of
\citet{wright} and \citet{taylerpol}. Using coordinates adapted to the
magnetic-field geometry, these authors found
unstable perturbations in the vicinity of the `neutral line', where
the magnetic field vanishes; see figure \ref{fieldlines}. Instability
was predicted to occur after approximately one Alfv\'en
crossing time --- the time taken for a magnetic perturbation to travel
around the star.

The existence of unstable perturbations around the neutral line of a
poloidal field is
analogous to the \emph{toroidal}-field case, for which \citet{taylertor}
showed the existence of an $m=1$ instability localised around the
magnetic field's symmetry axis (the region where a toroidal field
vanishes). A more general result for poloidal-field stability was
given by \citet{van_assche}, who showed that any poloidal field with
closed field lines is unstable around the neutral
line. \citet{taylerpol} suggest that the open field lines, in
contrast, may help \emph{reduce} the effect of the instability.

In contrast to the toroidal-field case, there does not seem to be a
definite conclusion from these analytic studies about which values of
azimuthal index $m$ correspond to the most unstable modes; this is a
consequence of them utilising field-line adapted coordinates rather than
global spherical polars. One might, then, expect instabilities to be
present for a variety of $m$ --- and this was indeed found in the
numerical studies of \citet{gepp_rhein} and \citet{braithpol}.

Analytic work \citep{pitts_tayler} suggests that rotation may help to
stabilise magnetic fields; this conclusion was backed up by the
results of numerical evolutions reported in \citet{gepp_rhein}. In
contrast, the work of 
\citet{braithpol} finds that rotation plays no stabilising role on
poloidal fields; however, since this study models a star as spherical
(with no distorting centrifugal force) it may be less relevant for
describing fast-rotating, highly distorted stars.

To summarise, the signatures of the poloidal-field instability
described in earlier work are: localised unstable growth around the
neutral line; onset after one Alfv\'en crossing time; (probable)
existence for a variety of $m$; (perhaps) stabilised by rotation. With
these in mind, we now turn to our results for the behaviour of
perturbations of a star with a poloidal field.

\begin{figure}
\begin{minipage}[c]{0.48\linewidth}
\psfrag{emag}{$\delta\hat{M}$}
\psfrag{t}{$\hat{t}$}
\psfrag{low}{low}
\psfrag{medium}{medium}
\psfrag{high}{high}
\includegraphics[width=\linewidth]{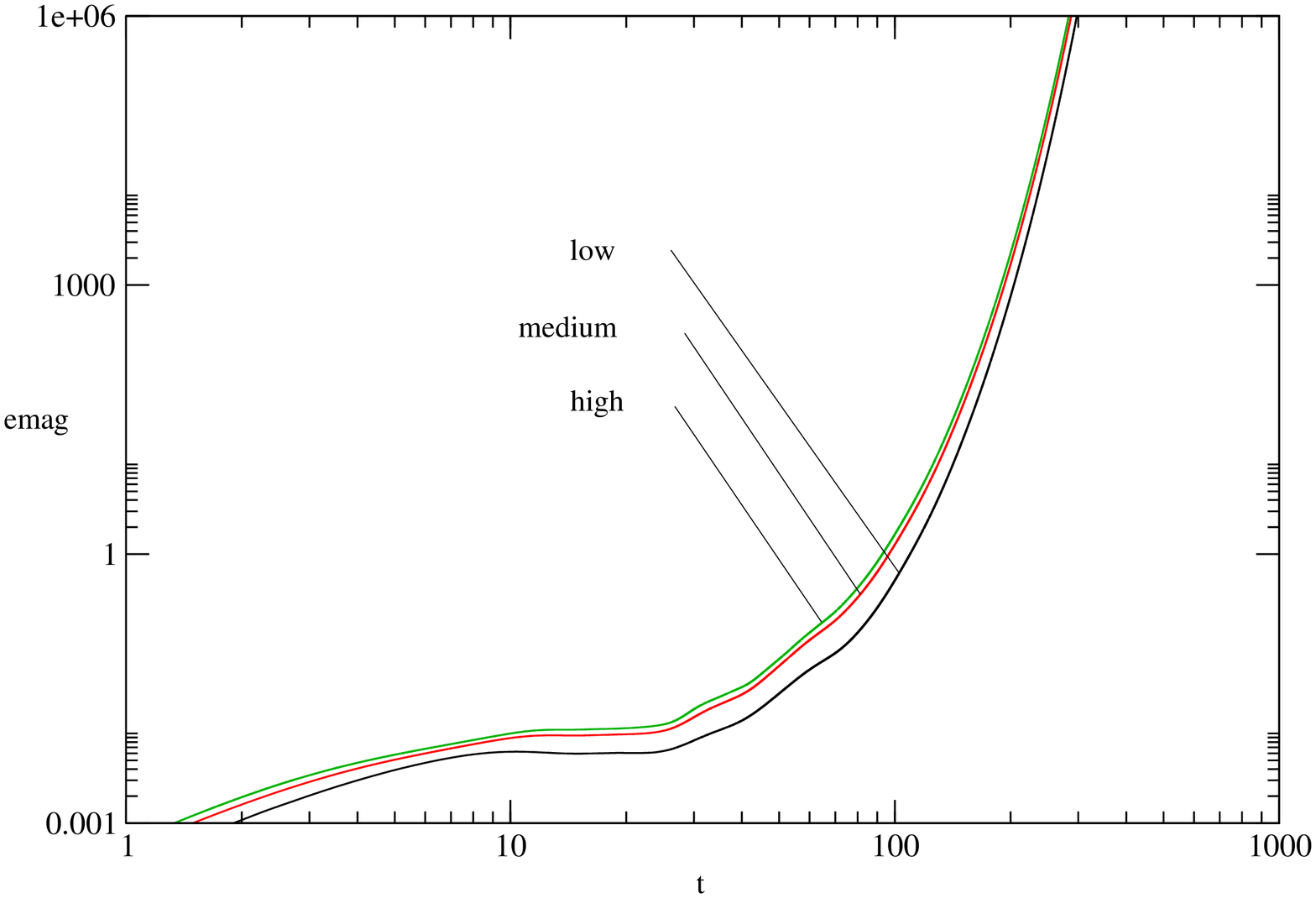}
\caption{\label{instab_norot}
         $m=2$ instability of a poloidal magnetic field in a
         nonrotating star. We plot the magnetic energy
         $\delta\hat{M}$ against time $\hat{t}$, both in
         dimensionless form, for three different grid resolutions. We
         see that the onset time for the instability is independent of
         resolution (appearing at around the expected value of
         $\hat{\tau}_A\approx 77$), and its growth rate converges,
         consistent with it being a physical instability. The
         results are for a star with average field strength
         $\Bav=3.0\times 10^{16}$ G.}
\end{minipage}
\hfill
\begin{minipage}[c]{0.48\linewidth}
\psfrag{emag}{$\delta\hat{M}$}
\psfrag{t}{$\hat{t}$}
\psfrag{om0}{$\hat{\Omega}=0.0$}
\psfrag{om158}{$\hat{\Omega}=0.158\approx 0.22\Omega_K$}
\psfrag{om305}{$\hat{\Omega}=0.305\approx 0.42\Omega_K$}
\includegraphics[width=\linewidth]{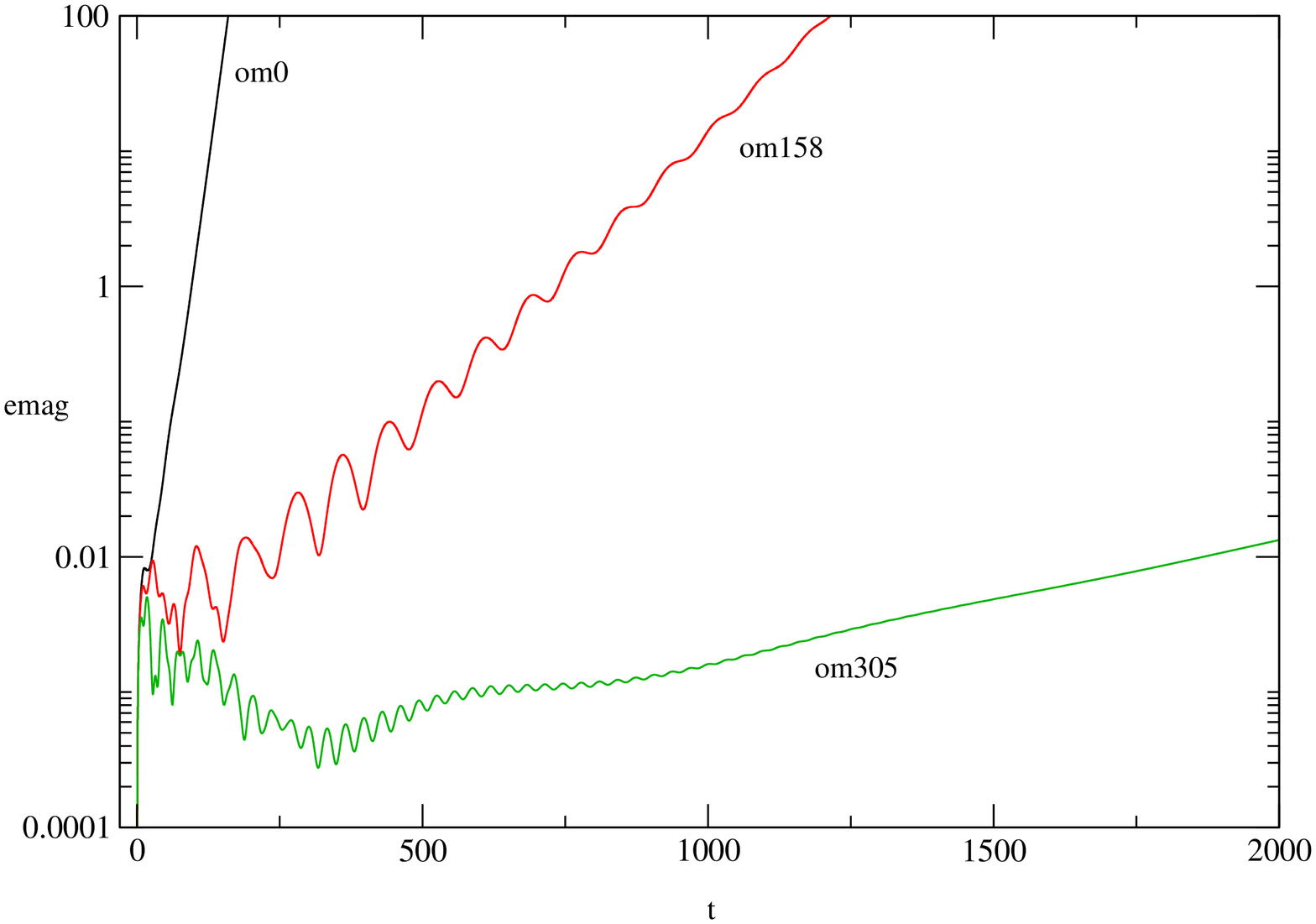}
\caption{\label{instab_rot}
         The stabilising effect of rotation on purely poloidal
         magnetic fields (for $m=2$ perturbations). The magnetic
         energy is plotted against time 
         for three different rotation rates. We see that increasing
         the rotation rate decreases the growth rate of the
         instability; i.e. the gradient of $\delta M$ is reduced in
         the regime where the instability dominates. As for the
         previous plot, each configuration has a field strength of
         $\Bav=3.0\times 10^{16}$ G.}
\end{minipage}
\end{figure}

To look for a poloidal-field instability, we begin by approximating
the Alfv\'en crossing time $\tau_A$ using volume-averaged background
quantities:
\beq
\tau_A \approx \frac{2R}{<\!c_A\!>}
          =     2R \sqrt{\frac{4\pi<\!\rho\!>}{\Bav^2}},
\eeq
where $R$ is the stellar radius, $c_A$ the Alfv\'en speed and angle
brackets denote volume averages. In dimensionless form, we find that
$\hat{\tau}_A\approx 77$ for a star with a (redimensionalised) field
strength of $\Bav=3.0\times 10^{16}$ G. In a plot of perturbed
magnetic energy $\delta M=\int(\delta B^2/8\pi)\ \rmd V$ against time, we
would therefore expect to 
see an instability manifest itself around $\hat{t}\approx 77$; this is
confirmed in figure \ref{instab_norot}. To confirm that the origin of
this instability is physical rather than numerical, figure
\ref{instab_norot} also shows that the result is similar for three
different grid resolutions: low (16 $r$-points $\times$ 15
$\theta$-points), medium ($32\times 30$) and high ($64\times 60$). The
three growth rates converge at second order, the intended accuracy of
the time-evolution code.

In figure \ref{instab_rot} we compare the behaviour of $\delta M$ for
stars with a fixed field strength ($3.0\times 10^{16}$ G) but
differing rotation rates, finding that rotation does act to slow the
instability's growth rate. We will return to quantify the change in
growth rate later in the section.

\begin{figure}
\begin{minipage}[t]{0.48\linewidth}
\begin{center}
\includegraphics[width=0.7\linewidth]{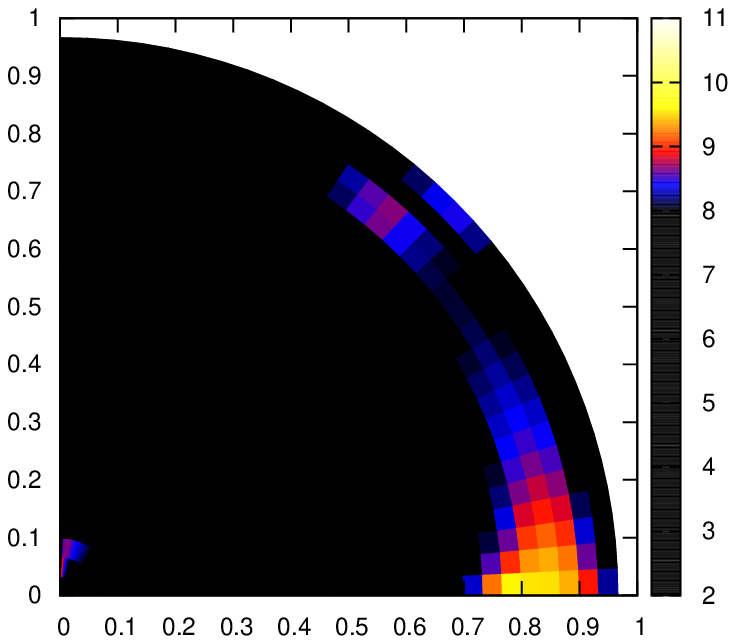}
\caption{\label{B_pert}
         The magnitude of the perturbed magnetic field after
         the onset of instability, plotted with a logarithmic
         scale. The most unstable perturbations are 
         visible around the neutral line, where the background field goes 
         to zero. This plot is for the $m=2$ instability, but very
         similar results emerge for $m=1$ and $m=4$ perturbations.}
\end{center}
\end{minipage}
\hfill
\begin{minipage}[t]{0.48\linewidth}
\begin{center}
\includegraphics[width=0.7\linewidth]{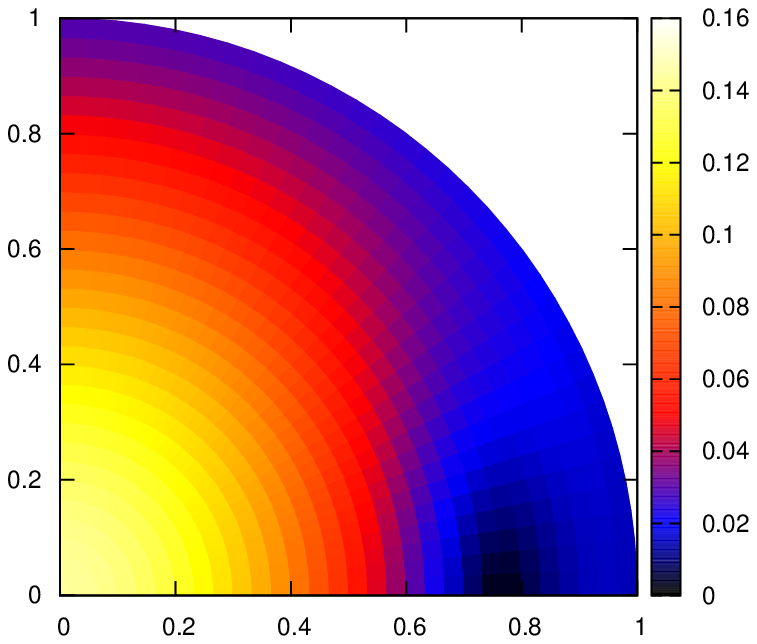}
\caption{\label{B_back}
         The poloidal field configuration of the background star. The
         field strength decreases to zero around the neutral line,
         which is located at a dimensionless radius $r/R \sim 0.8$ from
         the centre.}
\end{center}
\end{minipage}
\end{figure}

We next turn to the prediction that an instability should be localised
around the magnetic field's neutral line. From a global evolution, it
is not straightforward to separate the behaviour of the instability from 
stable perturbations. We attempt to isolate the unstable perturbation
by dividing the value of $|\dB|$ at each point by its value close to
the start of the evolution, in this way removing the shape of initial stable
perturbations. We plot the resulting (dimensionless) quantity
  with a logarithmic scale. After the onset of instability, all plots
we obtain are similar to that shown in figure \ref{B_pert}, for
azimuthal indices $m=1,2$ and $4$. Note that because our analysis
  is linear, the perturbed magnetic energy can grow indefinitely
  without diminishing the energy of the background configuration; in
  reality, nonlinear effects would become important after the initial
  period of unstable growth.

Comparing the shape of the unstable perturbation with the background
field configuration --- shown in figure \ref{B_back} --- we see that
the instability is clearly dominant around the neutral
point. Furthermore, it appears to remain localised in this
closed-field line region, fitting with the suggestion of
\citet{taylerpol} that open field lines are likely to provide a
stabilising influence.

\begin{figure}
\begin{minipage}[c]{\linewidth}
\psfrag{growth}{$\zeta$ [s${}^{-1}$]}
\psfrag{Bav}{$\Bav$ [$10^{16}$ G]}
\psfrag{Omega}{$\hat\Omega$}
\psfrag{m=1}{$m=1$}
\psfrag{m=2}{$m=2$}
\psfrag{m=4}{$m=4$}
\includegraphics[width=\linewidth]{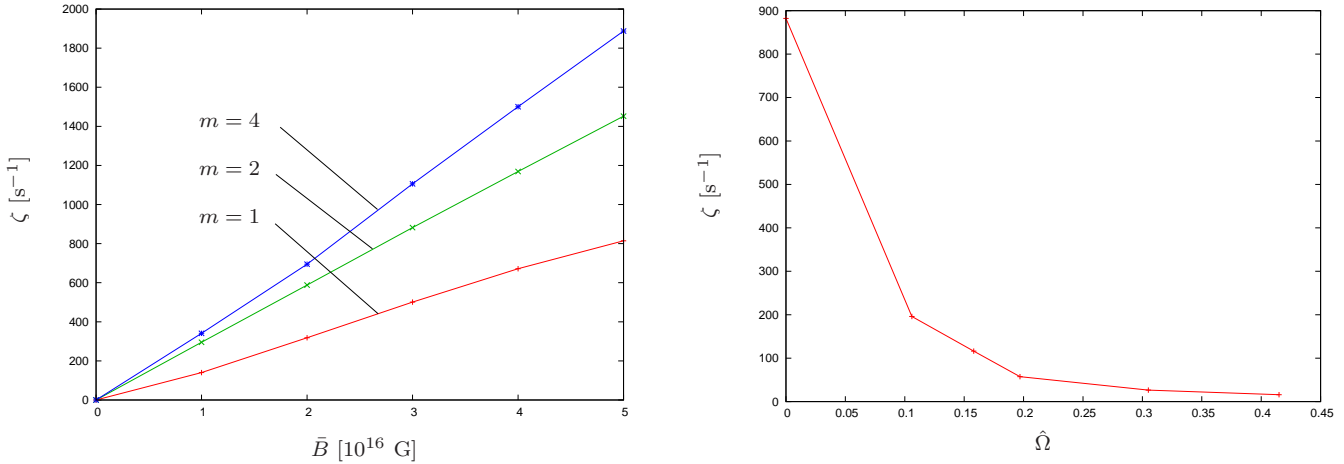}
\end{minipage}
\caption{\label{growthrates}
         Left: the instability growth rate (in seconds${}^{-1}$) plotted
         against field strength $\Bav$ for a series of nonrotating
         stars; we see that the dependence is linear for these field
         strengths. Right: the effect of rotation on the growth rate
         of the $m=2$ instability in a star with $\Bav=3.0\times
         10^{16}$. Note that the highest rotation rate shown is only $\sim 58\%$
         of Keplerian velocity; beyond this value the code is less
         (numerically) stable.}
\end{figure}

Having studied qualitative features of poloidal-field instabilities,
we conclude this section with some quantitative results. As a measure
of instability we define a growth rate $\zeta$ by
\beq
\zeta\equiv\frac{1}{\Delta t}\ \Delta\brac{\ln\brac{\frac{\delta M}{M_0}}},
\eeq
where $M_0$ is the magnetic energy of the background star. This simply
measures the degree of exponential growth of the unstable mode and is
easy to identify once the instability has come to dominate the
behaviour of the system. In figure \ref{growthrates} we use $\zeta$ to
investigate the dependence of the instability on the background field
strength and rotation rate of the star. From the left-hand plot we see
that the growth rate scales linearly with field strength, with
higher-$m$ instabilities appearing to grow faster. In the right-hand
plot of figure \ref{growthrates} we plot the $m=2$ instability growth
rate as a function of rotation frequency, up to $\Omega\approx
0.58\Omega_K$. Rotation is seen to reduce 
the growth of this instability, but with finite-duration evolutions we
cannot conclusively say that it will be removed altogether ---
however, we suggest that it may be stabilised in a real
star by a combination of rotation and some additional physics
(e.g. viscosity). A similar picture of rotational stabilisation seems
to emerge for stars with $\Omega>0.6\Omega_K$ and for $m=1$ and $m=4$
instabilities, although evolutions in these cases appear to be more
prone to late-time (numerical) instability.

As mentioned in sections \ref{BCs} and \ref{initial_data}, each
  evolution of perturbations we perform uses one of two possible
  symmetry classes, dictating the equatorial symmetry properties of each
  perturbation variable. Perturbations in the $\clP^+$ symmetry class
  evolve stably for many Alfv\'en timescales and we are able to study
  the corresponding `polar-led' modes (see the following section). The
  generic instability of poloidal fields --- explored in this section
  --- manifests itself only in the behaviour of $\clP^-$
  perturbations. We may understand this through a more detailed discussion
  of the study of \citet{taylerpol}. These authors work in a system of coordinates
  fitted to closed field lines, the geometry of which is a torus
 enclosing the neutral line. They first `open up' this torus into a
 cylinder to enable comparison with known instabilities in a
 cylindrical field geometry (see \citet{taylerpol}, figure 2).  In the
 simplest case there are two main instabilities: the varicose
 instability (also known as the sausage instability)
 and the kink instability. Within a star the most unstable
 perturbations will be non-radial --- those that do not involve work
 against the star's gravitational potential. Since the
 varicose instability \emph{does} involve radial motion, the kink
 instability is likely to be the dominant one in stars with poloidal
 fields. In the spherical polar coordinates of the star, the kink
 instability involves motion of fluid elements in the
 $\theta$-direction across the equatorial plane. Along this plane
 $v_\theta=0$ for the symmetry class $\clP^+$ but not for the class
 $\clP^-$ --- and so unstable behaviour is only seen in the latter
 symmetry class. We believe this is the first numerical confirmation of
 this aspect of poloidal-field instabilities.

\section{Oscillation modes}

\begin{figure}
\begin{minipage}[c]{\linewidth}
\begin{center}
\psfrag{Bav}{$\Bav$ [$10^{16}$ G]}
\psfrag{sigma}{$\hat\sigma$}
\psfrag{a1}{$a_1$}
\psfrag{a2}{$a_2$}
\psfrag{a3}{$a_3$}
\includegraphics[width=0.6\linewidth]{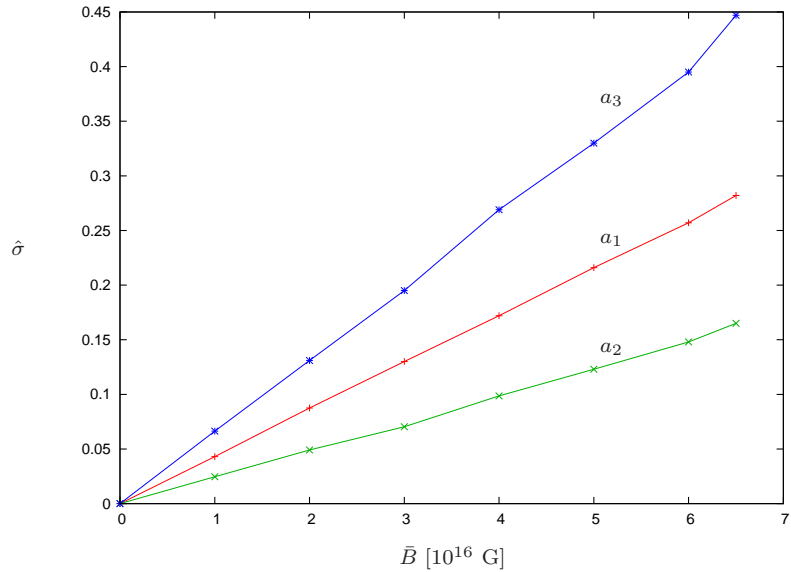}
\end{center}
\end{minipage}
\caption{\label{amodes_B}
         Polar-led $m=2$ Alfv\'en modes ($a$-modes) for a
         nonrotating star. We see that
         the mode frequency $\sigma$ varies linearly with $\Bav$ as
         expected, vanishing for unmagnetised stars (where the
         lowest-frequency mode is the $f$-mode). At the highest field
         strengths we see some possible deviation from the linear relation
         $\sigma\propto\Bav$. We are unable to produce a similar plot
         for axial-led $a$-modes, as these oscillations were
         swamped by the dominant unstable mode.}
\end{figure}

As described above, perturbations in the $\clP^-$ symmetry class are
dominated by unstable behaviour, preventing us from resolving this
class of Alfv\'en mode ($a$-mode) frequencies. In the $\clP^+$ class,
however, the perturbations behave in a stable oscillatory manner; in this
section we study their mode frequencies. All results reported in
this section are for $m=2$ 
modes --- both for brevity and because these are easiest to resolve
with our code. To check that these are representative of other
azimuthal indices we performed evolutions for $m=1$ and $m=4$
oscillations too; the results suggested that there are no qualitative
differences, but that $m=1$ modes are slightly lower-frequency and
$m=4$ slightly higher-frequency than the $m=2$ case.

Before looking at results for oscillation modes, we discuss some
terminology we will use to describe modes --- this is based on
notation employed by \citet{lock_fried} to describe inertial modes
($i$-modes). The eigenfunctions of
$i$-modes are --- with the exception of the $r$-mode --- a sum of 
spherical harmonic contributions $Y_{lm}$ and so cannot be labelled with a
single index $l$ (although they do have a single $m$). In
all cases, however, the sums are dominated by the $Y_{lm}$
contributions from $l=m$ up to some value $l=l_0$; the contributions
beyond $l_0$ are all far smaller. There may be several modes of the
same $m$ and $l_0$, so these are enumerated with an additional index
$k$ and denoted ${}^{l_0}_mi_k$. Since we focus on $m=2$ modes, we
will suppress the $m$-index. Finally, we recall from section
  \ref{BCs} that we will have two classes of oscillation, corresponding
  to the two symmetry classes: these are axial-led ($\clP^-$)
  and polar-led ($\clP^+$) modes. \citet{lock_fried} showed that
  for axial-led modes the lowest contributing $Y_{lm}$ to their
  eigenfunction (i.e. $l=m$) is axial and for polar-led modes the
  lowest term is polar.

\subsection{Modes of a nonrotating magnetised star}

In the nonrotating case, we find three peaks at
lower frequency than the $f$-mode in the frequency spectrum; we
identify these as magnetically-restored modes. We label these modes
$a_1,a_2$ and $a_3$ based on the amplitude of their peaks, from
strongest to weakest. We expect $a$-mode frequencies to be proportional
to the Alfv\'en speed $c_A=B/\sqrt{4\pi\rho}$ and hence scale roughly
linearly with $\Bav$; this is borne out in our results, shown in
figure \ref{amodes_B}.

We can gain some understanding about the eigenfunction structure of
these $a$-modes by comparison with the behaviour of the code for
inertial modes (in unmagnetised stars). In this case we find that the
lowest-$l_0$ modes have the highest-amplitude peaks in frequency space ---
these modes are excited more strongly because of the finite resolution of
the numerical grid, combined with the low-$l$ initial data we
use. This provides us with a useful rough diagnostic to identify
$a$-modes: we suggest that the eigenfunction of the strong peak $a_1$
contains lower $Y_{lm}$ contributions than the $a_2$ or $a_3$ modes.

In a perfectly conducting medium, like the model NS considered in this
paper, magnetically-restored oscillations can occur in a continuous
band of frequencies rather than being discrete global modes. This
result was established by analytic work for an incompressible medium
(see, e.g., \citet{goos_cont}) and more recent numerical work has
suggested that the axisymmetric oscillations of compressible stars may
form a continuum too
\citep{sotani_ax,cerda,colaiuda}. It is known, however, that
dissipative effects like viscosity and resistivity can act to remove the
continuum (e.g. \citet{ireland}).

To test for a mode continuum, we look at the oscillation frequencies
of perturbed quantities at different points within the star. If our
system has discrete global modes we expect all these local oscillation
frequencies to be equal; if there is a continuous mode spectrum then
oscillation frequencies will be position-dependent. From our
evolutions we find the former: mode frequencies at different points
within the star are equal, within the resolution dictated by the
length of our evolutions (of the order 1\%). Whilst this appears to
contradict recent studies on 
magnetar QPOs that \emph{have} found evidence for a mode continuum
(among them \citet{sotani_ax}, \citet{colaiuda} and \citet{cerda}), it
should be emphasised 
that these studies are not quite comparable. For one, we study
non-axisymmetric oscillations, rather than the $m=0$ modes in these
earlier studies; in addition we have only looked at polar-led
modes, in contrast with the axial oscillations of the other
papers. Our work is, however, in agreement with the study of polar
Alfv\'en modes by \citet{sotani_pol}, who also found a discrete
oscillation spectrum.

\subsection{Modes of a rotating magnetised star}

\begin{figure}
\begin{minipage}[c]{\linewidth}
\begin{center}
\psfrag{Om}{$\hat\Omega$}
\psfrag{sigma}{$\hat\sigma$}
\psfrag{a1}{$a_1$}
\psfrag{a2}{$a_2$}
\psfrag{a3}{$a_3$}
\psfrag{3i1}{${}^3i_1$}
\psfrag{3i2}{${}^3i_2$}
\psfrag{5i1}{${}^5i_1$}
\psfrag{5i2}{${}^5i_2$}
\psfrag{5i3}{${}^5i_3$}
\psfrag{5i4}{${}^5i_4$}
\includegraphics[width=0.8\linewidth]{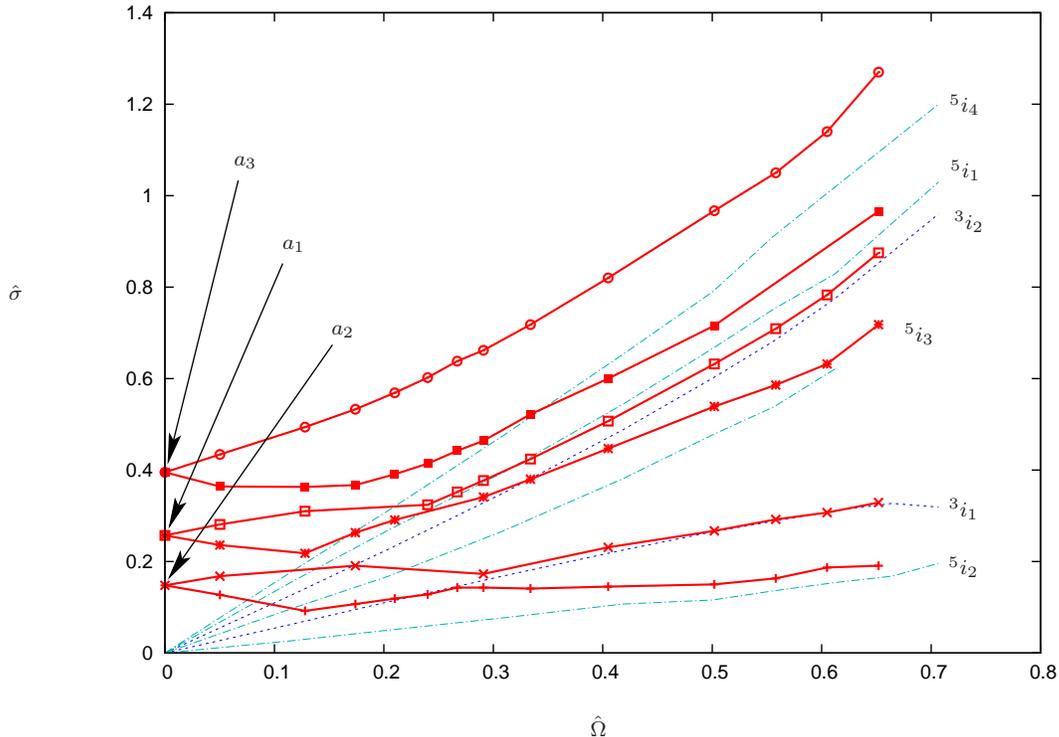}
\end{center}
\end{minipage}
\caption{\label{amodes_rot}
         Polar-led $m=2$ hybrid magneto-inertial modes, for a
         neutron star with
         field strength $\Bav=6.0\times 10^{16}$ G; the plot is clearer
         for this highly magnetised star than at lower values of
         $\Bav$. When $\Omega=0$ there are three pure Alfv\'en modes:
         $a_1,a_2$ and $a_3$. Rotation splits each into co- and
         counter-rotating modes, which are seen to become known
         inertial modes for high rotation rates (with the possible
         exception of the highest-frequency mode; see text), although the large
         number of avoided crossings makes it difficult to track each
         mode individually. In these dimensionless units Keplerian
         velocity $\hat{\Omega}_K\approx 0.72$.}
\end{figure}

Next we consider the mode spectrum of a rotating star with a poloidal
magnetic field, but let us first recall our results on toroidal-field
oscillation modes from \citet{tor_mode}. In this earlier work we found hybrid
magneto-inertial modes, whose character was Alfv\'en-like for slow
rotation and inertial-like in more rapidly-rotating stars
\citep{tor_mode}, so that their character depended on the ratio of
magnetic to kinetic energy $M/T$ --- we expect to see this happen in the
poloidal-field case too. In the toroidal-field case, we found two
polar-led $a$-modes in the nonrotating case. After rotational
splitting, one half of each $a$-mode appeared to become a
zero-frequency mode in the $M/T\to 0$ limit, whilst the other half of
each mode became an inertial mode: ${}^3i_1$ and ${}^3i_2$. This led
us to classify the $a$-modes as analogues of their respective inertial
modes, viz. ${}^3a_1$ and ${}^3a_2$.

In some respects, we find that polar-led oscillations of a poloidal-field
star appear to differ qualitatively from the results for toroidal
fields. In the non-rotating toroidal-field case we found two clear
$a$-mode peaks, whereas with a poloidal field we find three --- one of
which is stronger than the others. Adding rotation splits these three
modes into six --- i.e. three corotating and three counter-rotating
magneto-inertial modes; see figure \ref{amodes_rot}. Close to
Keplerian velocity $\Omega_K$ these all appear to become known
polar-led inertial modes: the two $l_0=3$ modes and the four $l_0=5$ modes
of a rotating unmagnetised star (see, e.g. \citet{lock_fried}).

The large number of avoided crossings (at which the character of two
modes changes) present in figure
\ref{amodes_rot} make it difficult for us to make conclusive
statements about the eigenfunctions of $a_1,a_2$ and $a_3$, because
unlike the toroidal case we cannot easily track these modes from the
Alfv\'en-dominated to inertial-dominated regimes. However, since each
$a$-mode splits into a co- and counter-rotating branch we expect them
to become a corresponding pair of co- and counter-rotating $i$-modes. We
also know from \citet{lock_fried} that ${}^3i_1,{}^5i_1,{}^5i_2$
corotate with the star and ${}^3i_2,{}^5i_3,{}^5i_4$ are
counter-rotating. Combining this information with the speculation that
the strong peak $a_1$ is a lower-$l_0$ mode than $a_2$ and $a_3$, we
suggest that $a_1$ becomes the pair of modes 
$\{{}^3i_1,{}^3i_2\}$ as $M/T\to 0$, whilst
$a_2\to\{{}^5i_2,{}^5i_3\}$ and $a_3\to\{{}^5i_1,{}^5i_4\}$.

Finally, we note the slightly anomalous behaviour of the
highest-frequency hybrid mode shown in figure \ref{amodes_rot}, which
begins at $\Omega=0$ as one branch of the $a_3$-mode. This mode does
not approach its apparent inertial counterpart ${}^5i_4$
as closely as the other hybrid modes in the high-$\Omega$ regime. Since
the other five hybrid modes \emph{do} convincingly become inertial
modes, however, we suggest that this discrepancy is due to an avoided
crossing with the corotating branch of the $f$-mode (not shown in the
figure) at $\hat\Omega\approx 0.6$.

\section{Poloidal-field effects in magnetars and pulsars}

In the previous two sections we presented results for instabilities
and oscillation mode frequencies at a variety of field strengths and
rotation rates. Here we will apply these to two specific cases: a
canonical model magnetar with\footnote{A typical rotation frequency
  for magnetars is $\nu=0.1$ Hz (equivalently $\hat\Omega\approx
  5\times 10^{-5}$ in dimensionless units), but we will assume 
  that such a small value is negligible when $\Bav\sim 10^{16}$ G
  --- as suggested by figures \ref{growthrates} (right-hand plot) and \ref{amodes_rot}.}
$\nu=0$ Hz and $\Bav=10^{16}$ G and a 
model pulsar-NS with $\nu=1$ Hz (a common, if slightly low value)
and $\Bav=10^{14}$ G (assuming a fairly high surface field strength of
$10^{13}$ G); in both cases we assume that the volume-averaged
magnetic field $\Bav$ is an order of magnitude greater than the
surface field strength. 

We first consider the stability of poloidal fields in each model
star. Although the magnetar field strength is 100 times that of the
pulsar, a purely poloidal field would be unstable in both --- however
the instability growth rate, linear in $\Bav$, would be 100 times slower in the
pulsar, even if it were nonrotating. The additional effect of its
rotation will further slow the growth of the pulsar's poloidal-field
instability. For the magnetar we see from the left-hand plot of figure
\ref{growthrates} that its growth rate $\zeta\approx 300$
s${}^{-1}$; equivalently its $e$-folding timescale is around 3
ms. If the pulsar were nonrotating its $e$-folding timescale would be
0.3 s; rotation will increase this somewhat, though we cannot quantify
this using our results. This is because our
rotating background stars are constructed by specifying the oblateness
of the star (which depends on the grid spacing) rather than its
rotation frequency, so our slowest-rotating stellar models still
rotate far faster than the 1 Hz of the model pulsar we wish to discuss
here.

Although a purely poloidal field is unstable for a number of azimuthal
indices, the instability is highly localised; this was predicted by
\citet{taylerpol} and our results (figures \ref{B_pert} and
\ref{B_back}) are in good agreement with this. As suggested by
\citet{wright}, a toroidal field which is fairly strong in the
neighbourhood of the neutral line could remove any poloidal-field
instabilities. Interesting, the toroidal field would not need to be
\emph{globally} strong to remove such instabilities.
Based on this, we suggest that in a stable 
mixed-field configuration the maximum values of the poloidal and
toroidal components may well be comparable but their respective energies
need not be (the toroidal-field energy is likely to be considerably
smaller, since the toroidal component need only occupy a small volume of the
star). This suggests that the `twisted-torus' magnetic-field
configurations discussed by (for example) \citet{YYE},
\citet{landerjones} and \citet{ciolfi} may be stable equilibrium
solutions, despite having toroidal components whose energy is
$\lesssim 10\%$ of the total magnetic energy.

Since purely poloidal fields are generically unstable they are not
candidates for long-lived magnetic field configurations in
magnetars. For this reason, we cannot be confident that our results
for polar-led Alfv\'en oscillations should closely resemble those
of a real magnetar. In addition, it is unclear how many of the
observed magnetar QPOs (see \citet{watts_stroh} for details of
these) originate as Alfv\'en modes of the interior, as opposed to
elastic modes of the crust (for example). However, if future
observations of magnetar oscillations do include frequencies best
fitted by our results, it could indicate that these QPOs represent
$a$-modes of dominantly poloidal-field configurations.

The frequencies of the three polar-led $a$-modes $a_1,a_2$ and $a_3$ exist
in the ratio 7:4:11 (with a maximum discrepancy of 2\%). For our model
magnetar we would expect a strong Alfv\'en QPO (corresponding to
$a_1$) at 83 Hz, with less significant QPOs at 47 Hz ($a_2$) and 130
Hz ($a_3$). It is interesting to note that our predicted value of 83
Hz is rather close to the two dominant magnetar QPO frequencies
observed to date: 84 Hz for SGR 1900+14 and 92 Hz for SGR 1806-20
\citep{watts_stroh}.

For our model pulsar we know that the oscillations will be
hybrid magneto-inertial modes, but predicting whether they will be
more Alfv\'en-like or inertial-like is difficult --- as before, this
is because of finite resolution dictating the allowed rotation rates
for our background stars. To attempt to gauge the relative influences
of $\Omega$ and 
$\Bav$ on the model pulsar's oscillation modes, let us return to
figure \ref{amodes_rot} (for a $6\times 10^{16}$ G star). In this
figure the Alfv\'en mode $a_1$ is seen to split into two hybrid
modes, with frequencies $\sim 10\%$ different from their inertial
counterparts at $\hat\Omega=0.31$ ($\nu=600$ Hz). We also know that
hybrid-mode frequencies scale with $M/T$ (see \citet{tor_mode} for
more on this), and that
$M/T\sim\Bav^2/\Omega^2$. For our model pulsar $\Bav$ is a factor of
600 smaller than in the plot and $\Omega$ is also a factor of 600
smaller, so $M/T$ is roughly equal to its value for the figure
\ref{amodes_rot} star at $\hat\Omega=0.31$; from this we suggest
that the pulsar's oscillation modes will be dominantly inertial, with
a magnetic correction of $\sim 10\%$.

\section{Discussion}

In this paper we have studied the behaviour of perturbations of a
neutron star with a purely poloidal magnetic field. The background
equilibrium configurations are generated self-consistently; the
density distribution may be distorted by a combination of magnetic and
rotational forces and the magnetic field is multipolar and poloidal,
generalising earlier work for dipole fields with a simple analytic form.
 
A number of previous studies have shown that poloidal fields
generically suffer instabilities localised around the `neutral
line' (where the background magnetic field vanishes); we confirm
that these are also present in our global evolution of perturbations and
for our particular poloidal field. The instability is present for a
number of non-axisymmetric perturbations --- we have studied it
for azimuthal indices $m=1,2$ and 4 --- but only in one of the two
  symmetry classes of perturbation. We argue that
  this confirms a prediction from \citet{taylerpol} about which modes
  are most unstable. Our results suggest that rotation has a
stabilising effect on purely poloidal magnetic fields.

For the stable symmetry class we are able to evolve
perturbations for many Alfv\'en oscillation periods and hence extract
mode frequencies. These non-axisymmetric oscillations appear to be
global modes, in contrast with the continuous spectrum of axisymmetric
modes discussed recently by various authors
\citep{sotani_ax,cerda,colaiuda}. We 
find three Alfv\'en modes, whose frequencies scale linearly with the
field strength. These are rotationally split into pairs of modes, one
of which co-rotates with the star and one which is
counter-rotating. As the stellar rotation rate is increased (or the
field strength is reduced) these become inertial modes; so in the case
of a rotating magnetised star pure inertial modes and pure Alfv\'en
modes are replaced by hybrid magneto-inertial modes.

There are a few qualitative differences between the results presented
here and our earlier work on toroidal-field oscillations
\citep{tor_mode} and instabilities \citep{tor_instab}. Poloidal-field
instabilities are not predominantly confined to $m=1$ as
toroidal-field instabilities are, but instead exist for a variety of
$m$. The poloidal-instability growth rates we have studied ($m=1,2,4$)
are all lower than the $m=1$ toroidal-instability growth rate, though
of the same order of magnitude. It appears from our results that in
the poloidal-field case the instability is more localised than the
toroidal-field instability studied in \citet{tor_instab}. Perhaps the most
notable difference is that the poloidal-field instability only exists
in one symmetry class of perturbations rather than both. In terms of
oscillations, the character of hybrid magneto-inertial modes appears
different if the background field is poloidal rather than toroidal. In
the former case we find that each hybrid mode becomes inertial as
the rotation rate is increased; in the later case some of the hybrid
modes appeared to become zero-frequency modes in this limit.

Our conclusion that purely poloidal fields are generically unstable is
hardly surprising; however, it was not guaranteed that these localised
instabilities would show up in evolutions of a discretised set of
perturbations on a relatively coarse grid. Having shown that our code
(at fairly low resolution) is adequate to study these unstable
perturbations, we now discuss some other time evolutions of magnetar
oscillations which have \emph{not} found such instabilities and
speculate on reasons for this. The work of \citet{colaiuda}
uses a coordinate system adapted to magnetic field lines in order to
reduce the perturbation equations to a one-dimensional
system. In these coordinates there is no coupling between different
closed field lines, which together with a boundary condition of not
evolving perturbations on the neutral line allows for stable
evolutions without numerical viscosity. \citet{sotani_ax} did not employ
these field-line coordinates and had to use numerical viscosity to
suppress instabilities connected with evolving this
intrinsically 1D system in 2D. The
related work of \citet{cerda} does not mention instabilities and their
plots of local QPO amplitudes within the star show that there is
little or no growth immediately around the neutral line. Their work
uses the anelastic approximation to suppress fluid modes, but it is
not clear whether that should also remove poloidal-field instabilities.

All of these studies are based on long-term stable evolutions of
perturbations on a magnetised background star; furthermore, they
evolve axial perturbations, which our work indicates are particularly
unstable for poloidal fields. The important difference between their
work and ours appears to be that all three papers we have discussed
specialise to axisymmetric ($m=0$) 
perturbations, whereas we are currently only able to evolve
perturbations with $m\geq 1$. Axisymmetric perturbations may simply be
less unstable; evolving these perturbations with (for example)
numerical viscosity may be sufficient to remove instabilities
altogether.

We have discussed some of the roles a poloidal field may play in both
magnetars and pulsars. The unstable nature of purely poloidal
fields means that they are unlikely to be good models of any NS
magnetic field, although the instability growth rate is considerably
slower for pulsars than magnetars. We find that the dominant polar-led
Alfv\'en mode of a poloidal-field magnetar has a frequency of around
83 Hz, whilst pulsar oscillation modes will be hybrid magneto-inertial
in nature. Although dominantly inertial, these pulsar modes may have
magnetic corrections of $\sim 10\%$ for fairly typical field strengths
and rotation rates.

It has long been thought that a magnetic field may be stable with a
suitable combination of toroidal and poloidal components
(e.g. \citet{taylermix}). Since the poloidal-field instability only occurs
close to the neutral line, we suggest that the addition of a toroidal
component in this region alone may provide stabilisation. The toroidal
component would have to be locally comparable in strength with the
poloidal component, but need only occupy a small volume --- in this
case its contribution to the total magnetic energy could be quite
modest. From this we have an indication that some recent models of
mixed-field equilibria \citep{YYE,landerjones,ciolfi}
may be stable, despite having only a small percentage of magnetic energy
contained in the toroidal component. We hope to explore the stability
and oscillation spectra of these mixed-field configurations in a
future study.

\section*{Acknowledgments}

SKL acknowledges funding from a Mathematics Research Fellowship from
Southampton University. This work was supported by STFC through grant
number PP/E001025/1 and by CompStar, a Research Networking Programme
of the European Science Foundation. We thank Kostas Kokkotas, Kostas
Glampedakis and the anonymous referee for their helpful comments on
this work.

\label{lastpage}

\end{document}